\documentclass[runningheads]{llncs}
\usepackage{graphicx}
\usepackage{cite}
\usepackage{amsmath,amssymb,amsfonts}

\usepackage{algorithmic}
\usepackage{graphicx}
\usepackage{ntheorem}
\usepackage{textcomp}
\usepackage{bmpsize}
\usepackage{float}
\usepackage[table]{xcolor}
\usepackage{lipsum}
\usepackage{tabularx}
\usepackage{graphicx}
\usepackage{adjustbox}
\usepackage{subcaption}
\usepackage[colorlinks=true,urlcolor=black]{hyperref}
\usepackage[normalem]{ulem}
\usepackage{array}
\newcolumntype{L}[1]{>{\raggedright\let\newline\\\arraybackslash\hspace{0pt}}m{#1}}
\newcolumntype{C}[1]{>{\centering\let\newline\\\arraybackslash\hspace{0pt}}m{#1}}
\newcolumntype{R}[1]{>{\raggedleft\let\newline\\\arraybackslash\hspace{0pt}}m{#1}}
\newcolumntype{P}[1]{>{\centering\arraybackslash}p{#1}}
\usepackage{todonotes}

\usepackage{xspace}

\begin{document}
\title{Tracking Mixed Bitcoins}
\titlerunning{Tracking Mixed Bitcoins}

%
\author{Tin Tironsakkul\orcidID{0000-0000-0000-0000} \and
Manuel Maarek\orcidID{0000-0001-6233-6341} \and
Andrea Eross\orcidID{0000-0003-0899-7960} \and
Mike Just\orcidID{0000-0002-9669-5067}}

\authorrunning{T. Tironsakkul et al.}
%
\institute{Heriot-Watt University, Edinburgh, Scotland, UK}

\maketitle             
\begin{abstract}

Mixer services purportedly remove all connections between the input (deposited) Bitcoins and the output (withdrawn) mixed  Bitcoins, seemingly rendering taint analysis tracking ineffectual. In this paper, we introduce and explore 
a novel tracking strategy, called \emph{Address Taint Analysis}, that adapts from existing transaction-based taint analysis techniques for tracking Bitcoins that have passed through a mixer service. We also investigate the potential of combining address taint analysis with address clustering and backward tainting. 
We 
further
introduce a set of filtering criteria that reduce the number of false-positive results based on the characteristics of withdrawn transactions and evaluate our solution with verifiable mixing transactions of nine mixer services from previous reverse-engineering studies. Our findings show that it is possible to track the mixed Bitcoins from the deposited Bitcoins using address taint analysis and the number of potential transaction outputs can 
be significantly 
reduced with the filtering criteria.

 
 
\end{abstract}
%

%
\section{Introduction} \label{sec:intro}



A Bitcoin \emph{mixer service} (also commonly known as \emph{tumbler} or \emph{laundering service})
is a cryptocurrency service that 
allows
users to ``anonymise" their Bitcoins by eliminating any possible connection between their original deposited Bitcoins and the \emph{mixed} Bitcoins that they withdraw later from the service~\cite{Simon_bitter_2012, joancomart_anonymity_2015}. This mixing process
can make the tracking of Bitcoin movements between addresses challenging, such as when using techniques like \textit{taint analysis}~\cite{moser_inquiry_2013}.
Mixer services are also frequently used as one of the core components in transaction obscuring for illicit activities, such as theft, ransomware, and dark market trade~\cite{turner_bitcoin_2017, Samsudeen_illegal_2019}.



In a normal Bitcoin transaction, address $A$ would send Bitcoins directly to address $B$. However, this interaction establishes a connection between the two addresses in the blockchain, allowing anyone to observe the movement of Bitcoins~\cite{nakamoto_bitcoin:_nodate}. Mixer services attempt to prevent this traceability by serving as an intermediary between the two addresses where address $A$ deposits Bitcoins to a mixer service address (named the receiver address) for mixing purposes. Next, the mixer service uses another address(es) (named the delivery address) to deliver completely unrelated Bitcoins to address $B$ in withdrawn transactions.
%
%

As a result, the interaction between address $A$ and $B$ is obscured in the blockchain, 
as there is no direct connection or transaction between the two end-point addresses. 
Furthermore,
simple transaction tracking methods are incapable of tracking the actual exchange of Bitcoins between the two addresses. One 
method used to track the mixed Bitcoins is 
to calculate every possible combination on every transaction within the mixing time for the potential withdrawn transaction outputs~\cite{hong_demix_2018}, which requires significant computational resources.  
%

Few 
studies have investigated reverse-engineered mixer services 
to discover their mixing pattern~\cite{moser_inquiry_2013, balthasar_laundry_2017,wegberg_2018}. 
We are aware of only one study 
that 
proposed a tracking method for mixed Bitcoins, which adapted from the aforementioned approach,  
and evaluated their method on a single mixer service~\cite{hong_demix_2018}. In particular, we are not aware of any 
proposed
tracking method to overcome the transaction obscuring feature of mixer services. 

Hence, in this paper we 
introduce
a novel tracking method called \emph{address taint analysis} that focuses on tainting at the address level, whereas previous taint analysis approaches have focused on tainting at the transaction level.
%
%
We investigate this method, both on its own and in combination with other tracking methods such as \emph{address clustering} and \emph{backward tainting}.
We also introduce a set  of filtering  criteria in an attempt to reduce  the number of  false-positive  results, and we evaluate our solutions with verifiable mixing transactions of nine mixer services used in previous reverse-engineering studies.
The remainder of the paper is structured as follows. We describe the related work 
in Section~\ref{sec:background}. We define our new methods and filtering criteria in Section~\ref{sec:methodology}. Using the sample cases presented in Section~\ref{sec:data}, we evaluate the results of these methods and discuss the results in Section~\ref{sec:results}. In Section~\ref{sec:conclusion}, we conclude and discuss improvements we envision.


\section{Related Work}
\label{sec:background}

\subsection{Taint Analysis}
\label{sec:TaintAnalysis}

\emph{Taint analysis} is a transaction tracking method that determines the relationship or connection of addresses based on exchanges of specific Bitcoins in transactions~\cite{moser_inquiry_2013}. 
It is often adapted to track the movement of specific Bitcoins (e.g., stolen Bitcoins)
by classifying the tracked Bitcoins as tainted or clean and calculating the distribution of tainted Bitcoins used in 
subsequent 
transactions.


One taint analysis method, the \textit{Poison method}, considers all of the resulting transaction outputs as fully tainted~\cite{moser_towards_2014}. There are other tainting strategies that utilise different approaches of tracking and distributing Bitcoins, such as the \textit{Haircut method}~\cite{moser_towards_2014} and \textit{FIFO method} (First In, First Out)~\cite{ander_fifo_2018}. Taint analysis can also be performed backwards, where instead of tainting forward to the next transaction, the algorithm taints backwards following previous transactions, possibly all the way back to the coinbase transaction where the Bitcoins originate from~\cite{ander_fifo_2018}.

Aside from transaction tracking, taint analysis is utilised to measure the effectiveness of transaction obscuring methods where results with tainted connections indicate that the obscuring method is ineffective and can still be tracked~\cite{meiklejohn_privacy_2015, ruffing_shuffle_2017}. 
We hereafter refer to the original transaction-based taint analysis as \emph{transaction taint analysis} to distinguish it from the address-based taint analysis we define in this paper (see Section~\ref{sec:addr-taint}).



\subsection{Address Clustering}
\label{sec:addresscluster}

\emph{Address clustering} is a method that operates by grouping addresses into a cluster based on specific transaction behaviours. Address clustering methods are utilised in de-anonymisation which attempts to classify Bitcoin addresses likely to belong to the same user for tackling illegal activities~\cite{meikle_fist_2013}.


One address clustering method, called \emph{input-sharing clustering} or \emph{multi-input heuristic}, is based on the assumption that all the addresses that share inputs in the same transaction belong to the same entity because every input address must sign a digital signature with its private key for the transaction to be valid. As such, if there are two or more input addresses in the same transaction, these addresses are being controlled by the same user~\cite{harrigan_unreasonable_2016}.

Although this 
address clustering method has the advantage of relying only on information available inside the Bitcoin blockchain, it is frequently considered to be less effective for de-anonymisation purposes due to the existence of \emph{CoinJoin}\footnote{\emph{CoinJoin} is a method which allows multiple users to combine their Bitcoin transactions in a single transaction to improve their transaction privacy~\cite{Maurer_2017}.}. Input-sharing clustering classifies every address that shares input in the same transaction via CoinJoin as belonging to the same user. As a result, input-sharing clustering will create inaccurate address clusters that belong to different users and cannot be practically utilised for de-anonymisation purposes.

Another clustering approach, \textit{change address clustering},
operates by clustering the input addresses with the output addresses that are likely to be the change addresses -- this is an address that is owned by the transaction's sender and which receives the remaining change Bitcoins in the transaction\footnote{For example, if address $A$ uses a 10 Bitcoins input to send 5 Bitcoins to address $B$, the owner of address $A$ will need a change address (either address $A$ or another address) to receive 5 Bitcoins back.}~\cite{meikle_fist_2013}.


\section{Methodology}
\label{sec:methodology}


In this section, we describe the {address taint analysis} method and its combination with other tracking methods (address clustering and backward tainting). Subsequently, we discuss the filtering criteria we developed and the rationale behind them. 

To evaluate the effectiveness of our four tracking methods 
and filtering criteria, 
we compare the number of tainted transaction outputs of each method to the baseline of all outputs occurring in the same time frame. Our definition of the baseline is based on work from a previous study~\cite{hong_demix_2018}. 
%
%
%

\newtheorem*{baseline}{Baseline}
\begin{baseline} \label{baseline}
All outputs of every transaction recorded in the blockchain within the tainting time frame of a given sample case.
%
%
\end{baseline}

\subsection{Address Taint Analysis}
\label{sec:addr-taint}

The majority of mixer services usually utilise either a group of central addresses in order to combine and mix deposited Bitcoins from their users~\cite{moser_inquiry_2013, balthasar_laundry_2017, wegberg_2018}. We assume that the receiver and delivery addresses within the mixer services are both likely to interact with the central addresses at some point in time. 

\begin{figure}[ht]
  \centering
  \includegraphics[scale=0.45]{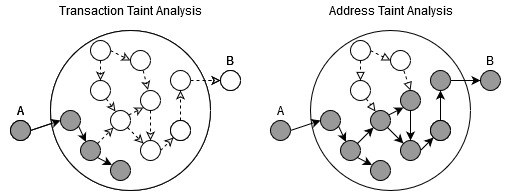}
  \caption{Transaction taint analysis and address taint analysis.}
  \begin{minipage}{.9\hsize}\footnotesize%
The figure depicts the difference between the transaction taint analysis and address taint analysis methods on an example mixing case that shows the deposited transaction from address $A$ and the withdrawn transaction to address $B$. A darker arrow and circle indicate an transaction and address that involves tainted Bitcoins, while a lighter one indicate it being clean.
\end{minipage}
  \label{taintcompare}
\end{figure}

Our 
taint analysis method, 
\emph{address taint analysis}, shown in Figure~\ref{taintcompare}, operates at the address connection level, where any address that receive Bitcoins from tainted addresses will be considered as a tainted address including every Bitcoin it possess at any point in time.
Existing taint analysis methods operate at the transaction level, where the tainted Bitcoins of 
a received address
do not affect other Bitcoins belonging to that address, unless 
they 
are used together in the same transactions.

The assumption for address taint analysis is that any transaction and address that can be connected to the receiver addresses at any point in time, whether directly or indirectly, may be related to the mixer service in some way. Therefore the objective of address taint analysis is not only to track the mixed Bitcoins, but also to map the network of address clusters and their transactions that may involve the mixer service operation, 
as in %
Bitcoin network analysis~\cite{Lischke2016AnalyzingTB}. Hence, address taint analysis tracking should be able to discover a relationship between the deposited and withdrawn transactions that the transaction taint analysis is unable to accomplish, as shown in Figure~\ref{taintcompare}. 

We describe three methods below for using address taint analysis (one further method, 
is described in Section~\ref{sec:backward-tainting}). The first method uses only address taint analysis. For the second and third methods, we investigate the potential of incorporating address clustering methods into the address taint analysis in order to improve the address cluster tracking results. As de-anonymisation is not our primary objective, we utilise the address clustering method to assist the address taint analysis algorithm for capturing relationships between addresses that are outside the scope of address taint analysis, which regard only the Bitcoin movement (address $A$ sends Bitcoins to address $B$).

\newtheorem{method}{Method}

\begin{method} \label{method:1}

Address taint analysis only.

\end{method}

The 
operation of address taint analysis used in this paper is conceptually similar to 
Poison transaction tainting~\cite{moser_towards_2014}
as the entire address is considered tainted, regardless of the number of tainted Bitcoins involved but goes further by affecting every Bitcoin possessed by the address throughout time. Since our main priority is to discover the connection between the deposited and withdrawn Bitcoins, other transaction tainting methods, which generally emphasise the distribution of tainting proportions, would not provide further information for this purpose.

As mixer services typically perform the mixing operation continuously, it is possible for the service to deliver Bitcoins that are already mixed prior to the time of the deposited transactions. As such, address taint analysis will also need to taint from the time period before the deposited transactions occurred. To put it simply, address taint analysis will taint all Bitcoins that the tainted addresses send both before and after the deposited transactions time.

\begin{method} \label{method:2}

Address taint analysis with input-sharing clustering.

\end{method}
%

We 
use
the input-sharing clustering method coupled with address taint analysis to taint any address that shares inputs with the tainted addresses. We use the same hypothesis as the original input-sharing clustering for our adaptation -- any address
that shares input in the same transaction with any tainted address is also likely to be one of the mixer service addresses and will be classified as a tainted address. 
%





\begin{method}\label{method:3}

Address taint analysis with input-sharing and output-sharing clustering.

\end{method}

As an augmentation to Method~\ref{method:2}, here we also
incorporate the output-sharing clustering method with the assumption that in the case of the mixing operation, the central addresses would often distribute the mixed Bitcoins to other mixer addresses first, before delivering them to the users. Consequently, we expect that output-sharing clustering should improve the chance of tracking such scenarios, even if the delivery addresses of the mixer service never send mixed Bitcoins to one another, or share input in the transaction. 

\subsection{Backward Address Taint Analysis}
\label{sec:backward-tainting}

\begin{figure}[ht]
  \centering
  \includegraphics[scale=0.45]{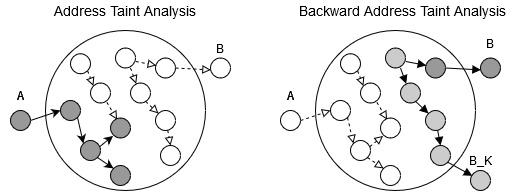}
  \caption{Address taint analysis and backward address taint analysis.}
\begin{minipage}{0.9\hsize}\footnotesize%
 The figure depicts a mixer service with two separate central groups tainted without and with backward address taint analysis. Notice the lack of any interaction between the address A and B groups. $B\_K$ represents the withdrawn output(s) from a known case used for backward address tainting in Method~\ref{method:4}. The lighter grey colour circles represent tainting results of backward address taint analysis and the darker grey colour circles represent tainting result of address taint analysis performed after backward address taint analysis.
 
\end{minipage}

  \label{cluster}
\end{figure}

The address taint analysis method operates with the assumption that the deposited and withdrawn mixer addresses may have a connection with each other via the central addresses, the analysis will not connect deposited inputs to the withdrawn outputs if there is no connection between the addresses involved, as shown in Figure~\ref{cluster}.
 
In such situation, address tainting from the deposited address cannot reach the withdrawal address. However, the knowledge of pre-existing withdrawal addresses could be used to identify the targeted withdrawal address. The search would consist of tainting backward from this known withdrawal address and then forward towards potential withdrawal addresses. 

Therefore, we introduce another method for this scenario by applying backward tainting to address taint analysis to create another tracking method called \emph{Backward Address Taint Analysis}. This method operates by tainting any address that sends Bitcoins to a tainted address. 
Rather than attempting to discover the connection between the mixed Bitcoins, the purpose of this method is to investigate whether it's possible to discover the address clusters used for withdrawn transactions. The idea would be that these addresses could subsequently be used to find the targeted withdrawn transaction outputs. Thus, this method operates in two steps, as described in the example below.

%
%



\begin{method} \label{method:4}
Perform Method~\ref{method:3} on the results of backward address taint analysis on the known pre-existing withdrawn transactions from the same mixer service.

\end{method}

Using the example from Figure~\ref{cluster}, Method~\ref{method:4} starts by performing the backward address taint analysis variation from the withdrawn transactions of a case 
from the same mixer service ($B\_K$) for three days to trace the mixed Bitcoins back to the central address clusters. Next, we use the results of the backward address taint analysis to perform address taint analysis at the time of the deposited transactions of the targeted sample case ($A$). 

\subsection{Filtering Criteria}\label{sec:refine}
\label{sec:withdrawal-props}

To 
further
reduce the number of false-positive results, we define five filtering criteria based on the information of the withdrawn transactions obtained from reverse-engineering experiments used in previous studies~\cite{moser_inquiry_2013, wegberg_2018, balthasar_laundry_2017}. 

The criteria 
can be applied for mixed Bitcoins in general with appropriate calibration. The calibration of the criteria parameters can also be specified to be stricter to reduce the false-positive results even further but this can increase the risk of missing the target. The parameters used in this experiment are obtained from observing the sample cases provided by the studies mentioned above. We set the parameters 
conservatively 
to reduce the risk of
losing the targeted withdrawn transactions for this experiment. 
In establishing the filtering criteria for our investigation we had the advantage of knowing the target withdrawn outputs that we were searching for. 
For future studies we plan to investigate the criteria on data with unknown target values. 


%
%

\newtheorem{criteria}{Criterion}

\begin{criteria}[Value of Withdrawn Bitcoins] \label{criteria:1}
The transaction output value of the targeted withdrawn transaction outputs cannot be higher than the deposited input value minus the mixing fee. 
\end{criteria}
%

As mixer services typically subtract a specific 
mixer service fee\footnote{Note that mixer service fee is different from Bitcoin transaction fee, which is mentioned in Criterion~\ref{criteria:5}.} from the initial deposited Bitcoins, the amount of the withdrawn Bitcoins would be lower than the original deposited amount. Depending on the mixer service, the mixing fee can vary in a specific range, 
such as between 1-2\% of the deposited Bitcoins. For this experiment, we use a minimum mixing fee for this criterion. 

This criterion does have at least one limitation, as it may be possible for the mixer services to combine the withdrawal of multiple deposited Bitcoins, which can make the withdrawn transaction outputs larger than the deposited input.

\begin{criteria}[Withdrawn Transaction's Shape] \label{criteria:2}
The number of transaction inputs and outputs of the targeted withdrawn transactions must be in the same pattern as the other withdrawn transactions by the same mixer service.
\end{criteria}


Reverse-engineering examples used in the literature~\cite{moser_inquiry_2013, balthasar_laundry_2017, wegberg_2018} show that the mixer services usually perform withdrawn transactions in a specific pattern. For example, one of the most common shapes of withdrawn transaction is in the form of a one-to-two addresses transaction where a single transaction output is sent to two addresses, one belonging to the user and the other to the mixer service. The number of transaction inputs and outputs of the targeted withdrawn transactions must be in the same pattern as the other withdrawn transactions by the same mixer service. 

A limitation of this criterion is that it is also possible for the mixer service to randomise the shape pattern or have an exception scenario (e.g., the withdrawn Bitcoins are large valued so that the service needs to combine other inputs in a withdrawn transaction) that can make the targeted withdrawn transaction different from the common pattern.

\begin{figure}[ht]
  \centering
  \includegraphics[width=5cm]{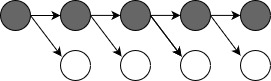}
  \caption{Example of a peeling chain.}
  \begin{minipage}{.9\hsize}\footnotesize%
  The figure depicts the peeling chain of a transaction chain that is commonly used by mixer services. A black circle represents a delivery address that belongs to the mixer service and a white circle represents a user address that receives a withdrawn transaction output.
  \end{minipage}
  %
  \label{peel}
\end{figure}

\begin{criteria}[Withdrawn Transaction Chain's Shape] \label{criteria:3}
If the mixing algorithm has a continuous withdrawn transaction chain pattern (e.g., peeling chain shown in Figure~\ref{peel}), either the transaction prior or after the targeted withdrawn transactions must have the same number of transaction inputs and outputs as the common pattern.
\end{criteria}


Following from Criterion~\ref{criteria:2},
the reverse-engineering results of the mixing sample cases indicate that multiple mixer services usually perform the withdrawn transactions in a continuous peeling chain, where a single transaction input with a large amount of Bitcoins is continuously peeled into two transaction outputs with one typically much smaller than the other~\cite{balthasar_laundry_2017}. 

As such, either the previous or next transaction of the targeted withdrawn transactions must also follow a similar pattern, accounting for the possibility that the targeted withdrawn transactions can be at the start or at the end of the withdrawn transaction chain.

Similar to a limitation for Criterion~\ref{criteria:2}, 
the mixer service can randomise the transaction chain shape or simply does not have one, which can increase the risk of missing the targeted outputs or make this criterion inapplicable. 


\begin{criteria}[Reused Input Address] \label{criteria:4}
The input address in the targeted withdrawn transactions is not used as transaction input more than once in its lifetime.
\end{criteria}

Our analysis of the verifiable mixing transactions from previous studies shows that the majority of the mixer services never reuse their delivery addresses before and after the withdrawn transaction.
Therefore, we can utilise this information as a criterion to exclude any transaction with input addresses that have been reused at any point in time. 

A limitation of this criterion is that although generally mixer services do avoid reusing the same address more than once, which is one of the most common Bitcoin privacy practices\cite{gaihre2018bitcoin},
it is possible for a mixer service to disregard this practice.


\begin{criteria}[Withdrawn Transaction Fee] \label{criteria:5}
The transaction fee value of the targeted withdrawn transactions must be the same as in other withdrawn transactions in the same time period.

\end{criteria}


From our own analysis of the verifiable mixing transactions from the previous studies, 
we also detect a specific pattern in the transaction fee values of the withdrawn transactions. In particular, the transaction fee values are of the same specific amount, such as 0.0005 BTC or 0.0001 BTC, even with a different transaction fee per byte ratio\footnote{Transaction fee per byte ratio is a metric to calculate the recommended transaction fee often used by miners to determine which transactions should be prioritised to maximise their mining profit.} and at a different time and day. This 
suggests
that mixer services generally do not automatically adjust the transaction fee setting in real-time but by a specific amount of time. As such, if the transaction fee always remains constant for the other withdrawn transactions in a similar time by the same mixer service, we can use the transaction fee as a criterion to exclude unrelated transactions.

Similar to the other criteria, it is possible for a mixer service to not have the practice of using a constant transaction fee for a period of time, which would make this criterion inapplicable.

\section{Sample Cases}
\label{sec:data}

We use 15 mixing transaction samples from previous studies~\cite{moser_inquiry_2013,balthasar_laundry_2017,wegberg_2018} 
which have shown that transaction taint analysis could not taint the withdrawn Bitcoins from the deposited Bitcoins. These studies perform reverse engineering on prominent mixer services shown in Table~\ref{tab:criteria}. 
%
%
These
studies perform reverse engineering on prominent mixer services: Blockchain.info's Shared Send function, Bitcoin Fog, Bitlaundry, BitLaunder, DarkLaunder, Alphabay and Helix Light. As one study~\cite{wegberg_2018} chose to not publicly name their tested mixer services, we exclude any identifiable information of the services and transactions, and refer to the mixer services from that study as "Unnamed".


For the address taint analysis experiment we use the transaction hash of deposited transactions to perform the address taint analysis and the transaction hash of withdrawn transactions are used to verify whether the address taint analysis can successfully reconnect the withdrawn Bitcoins back to the original deposited Bitcoins. If all of the targeted withdrawn transactions appear in the tainting results, we consider the experiment successful for that sample case.

In some of the mixing sample cases 
a \emph{change address} that belongs to the user and is reused to interact with the withdrawn Bitcoins later on
is used. This type of scenario 
can severely decrease the effectiveness of mixer services and make the mixed Bitcoins easily traceable. As user error is an extraneous variable that is not related to the mixer services and can affect the results of our experiment, we exclude any such change addresses from the deposited transactions. 

\section{Results and Discussion}
\label{sec:results}

\subsection{Address Taint Analysis}
\label{sec:res-addr-taint}

\begin{table}[ht]
\centering
\begin{tabular}{|rl|r|r|r|r|r|}
  \hline
  \multicolumn{2}{|l|}{\textbf{\textit{Case Service}}}  & \textbf{\textit{Baseline}} & \textbf{\textit{Method~\ref{method:1}}} & \textbf{\textit{Method~\ref{method:2}}} & \textbf{\textit{Method~\ref{method:3}}} & \textbf{\textit{Method~\ref{method:4}}} \\
\hline
  1 & Blockchain.info   &485,155 & --- & --- & \cellcolor{blue!23} 451,840  & n/a \\
\hline
  2 & Bitcoin Fog & 713,899 & --- &--- & --- & 682,901 \cellcolor{blue!26} \\ 
\hline
 3 & Bitcoin Fog & 1,525,276  & --- &--- & --- & 1,495,431 \cellcolor{blue!27}  \\  
 \hline
 4  & BitLaundry   & 1,013,374 & \cellcolor{blue!13} 841,563 & \cellcolor{blue!13}  845,846  & \cellcolor{blue!26} 976,745 & 989,264 \cellcolor{blue!27}  \\  
\hline
 5 & BitLaundry   & 1,016,043 & 843,277 \cellcolor{blue!13} & 847,662 \cellcolor{blue!13} & 984,089 \cellcolor{blue!27} & 463,584 \cellcolor{blue!5} \\ 
\hline
 6  & Unnamed 1 & 1,337,727 & 1,111,329 \cellcolor{blue!13} & 1,121,851 \cellcolor{blue!14} & 1,233,912 \cellcolor{blue!22} & n/a  \\
\hline
 7  & Unnamed 2 & 1,264,966 & 1,065,511 \cellcolor{blue!13} & \cellcolor{blue!14} 1,074,187 & 1,172,663 \cellcolor{blue!22} & n/a \\ 
\hline
  8 & Bitlaunder & 1,867,536 & \cellcolor{blue!10} 1,505,252 & \cellcolor{blue!10} 1,525,339 & 1,739,058 \cellcolor{blue!23} & 1,668,100 \cellcolor{blue!11} \\ 
 \hline
 9 & Bitlaunder & 2,156,487 & \cellcolor{blue!9} 1,712,671  & \cellcolor{blue!10} 1,731,779 & 2,006,007 \cellcolor{blue!23} & 1,919,906 \cellcolor{blue!19} \\ 
\hline
 10 & Darklaunder  & 1,712,521 & \cellcolor{blue!12} 1,407,023  & \cellcolor{blue!13}  1,421,073 & 1,606,730 \cellcolor{blue!24} &  1,637,210 \cellcolor{blue!25} \\ 
\hline
 11 & Darklaunder & 1,845,130 & \cellcolor{blue!10} 1,495,535 & \cellcolor{blue!13} 1,527,367 & 1,730,747 \cellcolor{blue!24} & 1,745,119 \cellcolor{blue!24} \\ 
\hline
 12 & Alphabay & 1,949,670 & \cellcolor{blue!11} 1,576,817  & \cellcolor{blue!13} 1,612,756 & 1,823,635 \cellcolor{blue!23} & 1,874,048 \cellcolor{blue!24}  \\  
\hline
 13 & Alphabay & 2,175,263 & \cellcolor{blue!11} 1,770,505  & \cellcolor{blue!13} 1,800,171 & 2,055,219 \cellcolor{blue!24} & 2,055,248 \cellcolor{blue!24} \\ 
\hline
 14 & Helix Light & 1,858,540 & \cellcolor{blue!5} 1,406,001 & \cellcolor{blue!7} 1,434,067 & 1,732,733 \cellcolor{blue!23} & 1,750,071 \cellcolor{blue!24}  \\ 
\hline
 15 & Helix Light & 1,777,542 & \cellcolor{blue!4} 1,326,101 & \cellcolor{blue!6} 1,358,043 & 1,669,601 \cellcolor{blue!24} & 1,669,599 \cellcolor{blue!24} \\ 
\hline
\end{tabular}

\caption{Address tainting results without filtering criteria.}
\begin{minipage}{.9\hsize}\footnotesize%
\footnotesize 


We indicate with --- that the method's experiment for the sample case was unsuccessful
and with n/a the absence of an experiment. (Method~\ref{method:4} requires another case of the same mixer service to performed).
The colour in each cell represents the percentage of the method's transaction output number compared to the Baseline method. The lighter the colour, the lower the percentage.
\end{minipage}
\label{tab:results}
\end{table}


As mixer services typically perform the mixing operation continuously, it is possible for the service to deliver Bitcoins that are already mixed prior to the time of the deposited transactions. We set the time limit for the address taint analysis operation to begin tainting from five days before the deposited transactions until the maximum amount of mixing time allowed by the mixer service (e.g., BitLaundry allows up to maximum 10 days mixing time). If the mixer service 
did
not have a mixing time setting, we 
set the time limit to three days.

As shown in Table~\ref{tab:refine}, the results of our experiment demonstrate that even 
mixed Bitcoins are not always perfectly immune to tracking. The majority of the sample cases show successful results overall except for the Blockchain.info and Bitcoin Fog cases where the target withdrawn outputs could not be found. 
The address taint analysis methods manage to accomplish the experiment's main objective, which is to reconnect the original deposited Bitcoins to the mixed Bitcoins, albeit with the extensive spreading of the tainted results. It should be noted that the number of transaction outputs in Table~\ref{tab:results} (and later in Table~\ref{tab:refine}) only count from when the deposited transactions occurred until the end of mixing time limit.

For the majority of sample cases, Method~\ref{method:1} yields the lowest number of transaction outputs compared to the other three methods and the Baseline method, followed by Method~\ref{method:2} and lastly Method~\ref{method:3}. The number of transaction outputs for Method~\ref{method:1} is considerably lower than those of the Baseline method at roughly 20\% for the successful cases. For example, Method~\ref{method:1} has 21\% (443,816) fewer transactions than the Baseline for case 9, and 17\% (171,811) fewer transactions for case 4.

The results of Method~\ref{method:2} are generally 
similar to
those of Method~\ref{method:1}. For example, 
Method~\ref{method:2} has only 1\% (8,676) more transactions than Method~\ref{method:1}
%
for case 7, 
and 2\% (35,939) 
more
for case 12. Meanwhile, Method~\ref{method:3} 
produces a greater
number of transaction outputs compared to the 
first two methods and 
is
much closer to the results of the Baseline method. For example, Method~\ref{method:3} has 12\% (199,707) 
more transaction output than Method~\ref{method:1} 
for case 10, 
and 6\% (105,791) 
fewer than the results of the Baseline method. As such, our 
results suggest 
that the incorporation of address clustering and backward tainting 
methods
is not always necessary for the tracking of mixer services though a few cases, Bitcoin.info's Shared Send and Bitcoin Fog are notable exceptions.

\subsection{Backward Address Tainting}
\label{sec:res-backward-tainting}

As shown in Table~\ref{tab:results}, the address taint analysis experiment on the Bitcoin Fog cases (2 and 3) produces unsuccessful results. This is because the mixer service keeps the deposited Bitcoins idle for as long as 6 months, which is outside the time period verification for our experiments.

While the initial deposited transaction for case 2 occurred on 29/04/2013, the deposited Bitcoins were not used at all until 07/11/2013, even though the withdrawn transactions occurred on 30/04/2013. This is similarly the situation for case 3. This type of scenario indicates that the central address clusters used for deposited and withdrawn transactions are separate and cannot be connected because of the time limit constraint in this experiment.

Method~\ref{method:4} shows successful results for all sample cases as shown in Table~\ref{tab:refine}. Although, aside from the Bitcoin Fog cases, the Method~\ref{method:4} tainting results (and the results after applying filtering criteria -- see Section~\ref{sec:res-withdrawal-props})
generally do not provide improved results compared to the other three methods. In particular, the number of transaction outputs resulting from Method~\ref{method:4} 
is higher than those of Method~\ref{method:3} for most cases. For example, 
Method~\ref{method:4} has 2\% (30,480) more transaction outputs than Method~\ref{method:3} for case 10
which is only 4\% (75,311)
lower than the Baseline method results. However, there are some exceptions where Method~\ref{method:4} performs better than Method~\ref{method:3} such as for cases 5 and 9,  where the number of transaction outputs are 53\% and 5\% lower than those of Method~\ref{method:3}, respectively.

Nevertheless, the results of the backward address tainting of Method~\ref{method:4} shows that it is possible to defeat the mixer service operation with separate central address clusters. If one can initiate the mixing transactions at the same time as the targeted mixing transactions so as to perform backward address tainting, one also can discover the central address clusters that are being used for the withdrawal of targeted mixed Bitcoins.

\subsection{Filtering Criteria}
\label{sec:res-withdrawal-props}

\begin{table}[ht]
\centering
\begin{tabular}{|c|c|c|c|c|c|c|c|}
  \hline
  \textbf{\textit{Service}} & \textbf{\textit{Criterion~\ref{criteria:1}}} & \textbf{\textit{Criterion~\ref{criteria:2}}}  & \textbf{\textit{Criterion~\ref{criteria:3}}}  & \textbf{\textit{Criterion~\ref{criteria:4}}} & \textbf{\textit{Criterion~\ref{criteria:5}}} \\
\hline
 Blockchain.info & 0.5\%  & one-to-one & one-to-two & Y & 10,000 Sat\\
\hline
 Bitcoin Fog &  1\% &  one-to-two & one-to-two & Y & 50,000 Sat \\
\hline
 BitLaundry  & 2.49\%  & one-to-two   & one-to-two & Y & 50,000 Sat \\
\hline
 Unnamed 1  & 1.5\% & one-to-two & one-to-two & Y & 10,000 Sat \\
\hline
 Unnamed 2  & 1\%  & one-to-two & one-to-two & Y & 10,000 Sat \\
\hline
 Bitlaunder & 2\% & N & N & Y & N\\
 \hline
 Darklaunder & 2\% & N & N & Y & N \\
\hline
 Alphabay & 10,000 Sat & one-to-two & N& N & N \\
\hline
 Helix Light & 2\% & one-to-many  & N & Y & 50,000 Sat \\
\hline
\end{tabular}
\caption{Sample mixer services and calibration of the filtering criteria.}
\begin{minipage}{.9\hsize}\footnotesize%
 The letter ``Y" indicates that the criteria can be applied to the mixer services and the letter ``N" indicates otherwise. The Bitcoin value is presented in Sat or Satoshis, which is the smallest unit of the Bitcoin (1 Bitcoin is equal to 100,000,000 Satoshis).
\end{minipage}
\label{tab:criteria}
\end{table}




After performing address taint analysis on each sample case, we applied the filtering criteria listed in Section~\ref{sec:refine} 
on each method's results for every case, as shown in Table~\ref{tab:criteria}.
%
%
%
While the majority of the sample mixer services employ a one-to-two peeling chain method (continuous one-to-two transaction), there are some exceptions. 
\begin{itemize}
\item The Blockchain.info's shared send function operates slightly differently than the other mixer services. Instead of peeling the withdrawn Bitcoins and sending them to the users directly, the service always peels off the withdrawn Bitcoins and transfers them to one of its addresses first, before sending them to the users in a one-to-one address transaction type. As such, Criteria~\ref{criteria:2} and~\ref{criteria:3} can still be applied for this mixer service case. 
\item The BitLaunder, DarkLaunder and Helix Light cases use a different version of a peeling technique. Instead of continuous one-to-two address transactions, the mixer services' algorithm 
peels a single large value transaction input to multiple transaction outputs (one-to-many). Additionally, the mixing algorithm of the BitLaunder and DarkLaunder cases do not always perform the withdrawal transactions in one specific pattern, hence we cannot apply Criteria~\ref{criteria:2} and~\ref{criteria:3} for these two mixer services' samples. Moreover, we also cannot apply transaction fee Criterion~\ref{criteria:5} as the mentioned mixer services regularly adjust the transaction fee based on the transaction size.
\end{itemize}

\begin{table}[ht]
\centering
\begin{tabular}{|rl|r|r|r|r|r|r|}
  \hline
  \multicolumn{2}{|l|}{\textbf{\textit{Case Service}}} &
  \textbf{\textit{Baseline}} & \textbf{\textit{Criteria}} & \textbf{\textit{Method~\ref{method:1}}} & \textbf{\textit{Method~\ref{method:2}}} & \textbf{\textit{Method~\ref{method:3}}} & \textbf{\textit{Method~\ref{method:4}}} \\
\hline
  1 & Blockchain.info & 485,155 & 87  & --- & --- & 84 \cellcolor{blue!27} & --- \\
\hline
  2  & Bitcoin Fog & 713,899 & 9,804 & --- & --- &--- &  9,428 \cellcolor{blue!26} \\
\hline
 3  & Bitcoin Fog & 1,525,276 & 12,945  & --- & --- & --- & 12,320 \cellcolor{blue!29}  \\
 \hline
 4 & BitLaundry & 1,013,374 & 24,885  & 23,617 \cellcolor{blue!25} & 23,661 \cellcolor{blue!25} & 24,696 \cellcolor{blue!29} & 24,710 \cellcolor{blue!29} \\ 
\hline
 5 & BitLaundry & 1,016,043 & 22,712 & 21,320 \cellcolor{blue!24}  & 21,361 \cellcolor{blue!24} &  22,376 \cellcolor{blue!28} &  10,236 \cellcolor{blue!1} \\
\hline
 6 & Unnamed 1 & 1,337,727 & 51,099 & 37,593 \cellcolor{blue!3}  &  37,951 \cellcolor{blue!4}  & 44,440 \cellcolor{blue!15}  & --- \\
\hline
 7 & Unnamed 2 & 1,264,966 & 48,626 &38,102 \cellcolor{blue!8}   &  38,587 \cellcolor{blue!9}  & 43,705 \cellcolor{blue!20}  & --- \\
\hline
 8 & Bitlaunder & 1,867,536 & 385,811 & 335,268 \cellcolor{blue!14}  & 335,966 \cellcolor{blue!15}  &  373,020 \cellcolor{blue!24}  & 347,389  \cellcolor{blue!20} \\
 \hline
 9 & Bitlaunder & 2,156,487 & 428,042 & 371,251 \cellcolor{blue!14}  & 372,103 \cellcolor{blue!15}  & 411,334 \cellcolor{blue!24}   & 380,805 \cellcolor{blue!18} \\
\hline
 10 & Darklaunder & 1,712,521 & 333,400 & 280,894 \cellcolor{blue!14}  & 288,290 \cellcolor{blue!14}  & 320,199 \cellcolor{blue!26}  & 319,447 \cellcolor{blue!25} \\
\hline
 11 & Darklaunder & 1,845,130 & 367,516 & 299,026 \cellcolor{blue!11}  & 315,474 \cellcolor{blue!15}  &  353,674 \cellcolor{blue!26}  & 354,285 \cellcolor{blue!26} \\
\hline
 12 & Alphabay & 1,949,670 & 181,512  & 154,426 \cellcolor{blue!15}  & 157,425 \cellcolor{blue!16}   & 174,487 \cellcolor{blue!26}  &  174,419 \cellcolor{blue!26}  \\
\hline
 13  & Alphabay & 2,175,263 & 227,718 & 178,960 \cellcolor{blue!12}  & 180,799 \cellcolor{blue!17}  & 215,534 \cellcolor{blue!24}  & 215,539 \cellcolor{blue!24}  \\
\hline
 14 & Helix Light & 1,858,540 & 6,329  & 5,764 \cellcolor{blue!21}  & 5,778 \cellcolor{blue!21}  & 6,166 \cellcolor{blue!27}  & 6,172 \cellcolor{blue!27}  \\
\hline
 15 & Helix Light & 1,777,542 & 6,160 & 5,731 \cellcolor{blue!23}  & 5,792 \cellcolor{blue!24}  & 6,009 \cellcolor{blue!27}  &  6,009 \cellcolor{blue!27}  \\
\hline

\end{tabular}
\caption{Address tainting results with filtering criteria.}
\begin{minipage}{.9\hsize}\footnotesize%
The Criteria column refer to the number of transaction outputs of the Baseline method after applying the filtering criteria. Each method column is the result of the method after applying filtering criteria. The colour in each cell represents the percentage of the method's transaction output number after applying the filtering criteria comparing to the Criteria column. The darker colour means that percentage is closer to 100\% of the baseline results after applying the criteria.
\end{minipage}
\label{tab:refine}
\end{table}

The address tainting results show significant improvement in terms of the number of transaction outputs for all of the methods including the Baseline method after applying the filtering criteria, as can be seen in the extensive reduction in the 
transaction outputs number shown in Table~\ref{tab:refine}. Assuming that our assumptions are correct, and the filtering criteria are correctly adjusted, this would mean that we've reduced a number of false positive transaction outputs. 

For the sample cases for which we can apply more filtering criteria, namely Cases 1 to 7, 14 and 15,
the number of false-positive transaction outputs is reduced by 90\% to as high as 99\%. However, the transaction output number after applying filtering criteria for the first three methods is closer to the Baseline method outputs at around 10\% lower.
For example, the number of transaction outputs for Method~\ref{method:1} for case 5 is reduced by 97\% (821,957), but when comparing to the Baseline method's results, the difference in transaction output number is becoming less after applying the filtering criteria from 17\% to only 6\% lower. 


While the sample cases that have less applicable filtering criteria, which are Cases 8 to 13, generally have lower reduction number of transaction outputs at around 80\%. When compared to the results of the Baseline method after applying filtering criteria, the number of transaction outputs show an increased reduction than for the other cases at around 20\% lower. For example, the number of transaction outputs for Method~\ref{method:1} for case 11 is reduced by 80\% (1,196,509) after applying the filtering criteria, but is 18\% (27,086) lower than the Baseline method.

However, there are cases where the results yield different result patterns. For example, for cases 7 and 8, the number of transaction outputs is much lower for the three methods compared to those of the baseline method, unlike the other cases with less applicable filtering criteria. Further, Method~\ref{method:1} for case 6 has a transaction outputs (after applying filtering criteria) that are 27\% less, and case 7 has 22\% less than the baseline. Interestingly, Helix light cases (case 14 and 15) show the largest reductions in the number of transaction outputs. We hypothesise this is due to the constant 50,000 Satoshis transaction fee used in the one-to-many transaction type that makes the withdrawn transactions extremely unusual compared to other transactions.


The differences in the results may be because the exploitable transaction patterns of mixer services have exceedingly unique patterns that make their transactions have characteristics that are considerably different from other transactions. Thus, this makes them less difficult to distinguish. We hypothesise that the fewer filtering criteria that can be applied to reduce the number of false-positive results, the more of an advantage the address taint analysis can provide over the Baseline method. Nevertheless, the significant reduction in transaction outputs suggests that the filtering criteria can be adopted for other tracking methods of mixer services in general.

\subsection{Limitations}

Despite the successful results and potential of the address taint analysis and filtering criteria, there 
are limitations of our approach that we discuss below.

The number of tainted transaction outputs with and without the filtering criteria are still relatively large when compared to the number of targeted withdrawn transaction outputs, as can be seen in Table~\ref{tab:results}. The address taint analysis in this paper taints the whole address, similar to the Poison method for transaction tainting, and does not utilise any other additional information besides the information of the deposited transactions.
Future research might attempt to further reduce the number of potential outputs.



Similar to other transaction tracking methods, address taint analysis can also be counteracted by the mixer services or the development of new privacy enhancement techniques that defeat the tainting algorithm. This is similar to how the CoinJoin method is introduced to oppose the input-sharing clustering method or mixer services to prevent transaction taint analysis tracking. Hence, the address taint analysis method will always require continuous development and improvement to remain applicable to new transaction obscuring techniques. 

The backward tainting approach is also not without challenges. The approach operates with the requirement that the attacker needs to know which mixer service is used for the targeted mixed Bitcoins. As shown in the Bitcoin Fog cases the receiver addresses are not reused addresses and have a very long idle time after receiving the deposited Bitcoins. It would be difficult to perform backward tainting within a similar time frame 
unless the attacker can identify the mixer service with other means in time, or simply perform backward tainting attacks on every mixer service that uses this type of mixing algorithm.  


Additionally, the risk increases if the mixer service uses a better randomised mixing algorithm to obscure any exploitable pattern. As the filtering criteria are currently designed based on the common transaction pattern found in the withdrawn transactions, the current criteria would be less effective as shown in the results of Table~\ref{tab:refine}. This ultimately has a high probability to create inaccurate results if the criteria are applied incorrectly. Thus, to avoid the risk of false incrimination of innocent users, the tracking method should always be utilised with caution and should only be implemented after a thorough exploration of the mixing algorithm involved.   

\section{Future Work and Conclusion}
\label{sec:conclusion}


As transaction obscuring methods improve, so should tracking methods to remain effective and relevant. We identify two possible improvements for both address taint analysis and filtering criteria, as follows;





\begin{itemize}
    \item \emph{Utilising external information to improve tracking results.}
The external information of address ownership can be collected from online websites, forums or services to exclude verifiable and reputable addresses that are unlikely to be a part of the mixer services likes cryptocurrency exchange service from the tainting results~\cite{fleder_bitcoin_2015}. This in turn can significantly reduce both the tainting operation time and the number of false-positive results. However, relying on external information comes with the risk of false or fabricated information depending on the source. Therefore, caution must be exercised to collect information from reliable sources.
\item
\emph{Incorporating more complex address clustering methods.}
There is another address clustering method that clusters based on transaction chain behaviour, instead of a single transaction~\cite{harrigan_unreasonable_2016}. For example, when one address distributes its Bitcoins to multiple other addresses, then those addresses transfer all of the distributed Bitcoins to a single address. We can assume that most of the addresses involved are likely to belong to the same user. Such a clustering technique could also be combined to address taint analysis similarly to the one we implemented in this paper.
\end{itemize}

The address taint analysis methods we propose in this paper have the
potential 
for
reconnecting the original deposited Bitcoins to the mixed Bitcoins -- this has not been possible with earlier taint analysis methods. We also illustrate that address taint analysis can be incorporated into other tracking methods such as address clustering and backward tainting methods for mixer services that utilise an irregular mixing algorithm. Although the number of false-positive results is still not substantially different between the Baseline and the other methods, by exploiting the transaction pattern of the withdrawn transactions to create filtering criteria, the number of false-positive results can be reduced further.

With further improvement, 
our approach could possibly
be used to assist cryptocurrency crime forensics in clearing the mystery of past illegal activities, such as exchange service thefts. 
Nevertheless, more mixing samples from other mixer services are still required for evaluating and improving the tracking method further, 
considering that 
mixer services are constantly evolving as new transaction obscuring techniques are introduced. 

\section*{Acknowledgment}


We would like to express our gratitude to Malte M\"oser, Rainer B\"ohme, Dominic Breuker, de Balthasar Thibault, Hernandez-Castro Julio C., Rolf van Wegberg, Jan-Jaap Oerlemans and Oskar van Deventer for providing with the transaction information of Bitcoin mixing samples used in their research work. Thanks also to Sasa Radomirovic who provided some feedback on earlier drafts of this paper. 

\bibliographystyle{splncs04}
\bibliography{mybibliography}
\end{document}